\documentclass[twoside,12pt]{article}  
\usepackage{ol2}
\usepackage{CJK}   
\usepackage{indentfirst}  
\usepackage{bm}    
\usepackage{graphicx}  
\usepackage{epsfig} 
\usepackage{amsmath}  
\usepackage{graphicx}
\graphicspath{{pict/}{./}}
\newcommand\pictc[5]{\begin{figure}
                       \centerline{\vspace{-1mm}
 \includegraphics[width=#1\columnwidth,height=0.7\textheight,keepaspectratio]{#3}}
                       \protect\caption{\protect\label{#4} #5}\vspace{-3mm}
                    \end{figure}            }

\newcommand\pict[4][1]{\pictc{#1}{!tb}{#2}{#3}{#4}}
\newcommand\rpict[1]{\ref{#1}}

\newcounter{Fig}

\newcommand{\be}{\begin{equation}}
\newcommand{\ee}{\end{equation}}

\begin{document}    




\begin{center}
\LARGE\bf Control of light scattering by nanoparticles with optically-induced magnetic responses$^{*}$  
\end{center}
\footnotetext{\hspace*{-.45cm}\footnotesize $^*$Project supported by the Australian Research Council Centre of Excellence for Ultrahigh Bandwidth Devices for Optical Systems (Project No. CE110001018) and the Future Fellowship (Project No. FT110100037).}
\footnotetext{\hspace*{-.45cm}\footnotesize $^\dag$Corresponding author. E-mail: ysk124@physics.anu.edu.au}

\begin{center}
\rm Wei Liu$^{\rm a,b)}$, \ \ Andrey E. Miroshnichenko$^{\rm b)}$, \ and  \ Yuri S. Kivshar $^{\rm b)\dagger}$
\end{center}

\begin{center}
\begin{footnotesize} \sl
${}^{\rm a)}$  College of Optoelectronic Science and Engineering, National University of Defence
Technology, Changsha, Hunan 410073, China\\

${}^{\rm b)}$ Nonlinear Physics Centre and Centre for Ultrahigh-bandwidth Devices for Optical Systems (CUDOS), Research
School of Physics and Engineering, Australian National University, Canberra, ACT 0200, Australia\\

\end{footnotesize}
\end{center}

\begin{center}
\footnotesize (Received X XX XXXX; revised manuscript received X XX XXXX)
\end{center}

\vspace*{2mm}

\begin{center}
\begin{minipage}{15.5cm}
\parindent 20pt\footnotesize
Conventional approaches to control and shape the scattering patterns of light generated by different nanostructures
are mostly based on engineering of their electric response due to the fact that most metallic nanostructures support
electric resonances in the optical frequency range. Recently, fuelled by the fast development in the fields of
metamaterials and plasmonics, artificial optically-induced magnetic responses have been demonstrated for
various nanostructures. This kind of response can be employed to provide an extra degree of freedom for the efficient control
and shaping of the scattering patterns of nanoparticles and nanoantennas. Here we review the recent progress in this research direction of nanoparticle scattering shaping and control through the interference of both electric and optically-induced magnetic responses. We discuss the magnetic resonances supported by various structures in different spectral regimes,
and then summarize the original results on the scattering shaping involving both electric and magnetic responses, based on the interference of both spectrally separated (with different resonant wavelengths) and overlapped dipoles (with the same resonant wavelength), and also other higher-order modes. Finally, we discuss the scattering control utilizing Fano resonances associated with the magnetic responses.
\end{minipage}
\end{center}

\begin{center}
\begin{minipage}{15.5cm}
\begin{minipage}[t]{2.3cm}{\bf Keywords:}\end{minipage}
\begin{minipage}[t]{13.1cm}
optically-induced magnetic response, scattering control, Mie resonance, Fano resonance.
\end{minipage}\par\vglue8pt
{\bf PACS: }
78.67.-n,   
42.25.Fx,   
73.20.Mf.   
\end{minipage}
\end{center}

\section{Introduction}

The seminal topic of light scattering by small particles has a long history, and it is of the fundamental importance for research in different areas of optical physics, including  sensing, solar cells, optical communications~\cite{Bohren1983_book}. Recently, with the rapid progress in nanoscience and nanotechnology, many types of nanoparticles have been fabricated being widely used in applications for biomedical labeling, impacting strongly other fields such as biology and medicine~\cite{Hirsch2003_PNAS, Atwater2010_NM, Huschka2011_JACS}. Stimulated by the flourishing fields of plasmonics and metamaterials and the renewed interest in the physics of nanoparticles, the study of light scattering by nanoparticles has gained unprecedented attention
during last years, and many novel scattering effects have been suggested and even demonstrated. This includes superscattering~\cite{Ruan2013_JPCC7324,Ruan2010_PRL}, clocking~\cite{Alu2009_PRL}, control of the scattering direction with metasurfaces~\cite{Yu2011_science,Ni2011_science}, artificial antiferromagnetism~\cite{Miroshnichenko2012_ACSNANO}, and nonlinear second-harmonic scattering~\cite{Zhang2011_NL}. The topic of light scattering by nanoparticles has also merged rapidly with the field of graphene~\cite{Geim2007_NM} and topological insulators~\cite{Hasan2010_RMP,Qi2011_RMP}, bringing out a lot of interesting phenomena~\cite{Gorodetski2010_PRB,Grigorenko2012_NP,Yin2013_science,Khanikaev2013_NM,Liu2013_thesis}.

 Among different research directions related to the physics of particle scattering, efficient and flexible scattering shaping is one of the most attractive topics. Conventional approaches for the scattering control and shaping rely solely on the engineering of the electric responses of nanostructures, due to the fact that most materials have only dominant electric responses, especially in the optical regime. The most widely employed response is the electric dipole (ED) resonance, the scattering pattern of which exhibits two typical features: (1) light is scattered equally to the backward and forward directions, and (2) orientation of the excited ED  is decided by the polarization of the incident wave, resulting in asymmetric azimuthal scattering patterns~\cite{Bohren1983_book}.  For various applications based on the particle scattering such as nanoantennas~\cite{ Curto2010_Science,Novotny2011_NP}, sensing~\cite{Kabashin2009_NM}, and photovoltaic devices~\cite{Atwater2010_NM}, the scattering patterns that are significantly different from that of a typical ED are usually required. One outstanding example is the requirement of the scattering pattern with suppressed backward scattering (reflection) and enhance forward scattering. The existing techniques usually rely on external additional coupled items, such as an extra reflector~\cite{Novotny2011_NP},  an extended substrate~\cite{Spinelli2012_NC}, Fabry-P\'{e}rot resonator like structures~\cite{Li2009_NL,Shegai2011_NL}, and/or other complicatedly engineered nanostructures~\cite{Aouani2011_NL,Rui2011_OL}, which could significantly hinder possible practical applications.

 It is known that introducing magnetic responses into scattering systems can bring an extra dimension for the scattering engineering~\cite{Bohren1983_book}. According to the symmetry of Maxwell's equations, if only magnetic response is present in the scattering structure, the scattering pattern would be the same as that of the corresponding electric response~\cite{Bohren1983_book}. However, if the electric and magnetic responses coexist and interfere with each other, then there will be an extra freedom and flexibility for an efficient scattering shaping~\cite{Love1976_RS,Kerker1983_JOSA,Nieto2010_OE,Garcia-Camara2010_OE}. The difficulty is that very few kinds of natural materials exhibit magnetic properties at optical frequencies. These are mainly due to magnetic dipole transitions in e.g. rare-earth atoms and are known to only operate in a narrow spectral range, and are, thus, usually highly lossy.

 Inspired by an earlier work on the optically-induced magnetic responses (OMRs) of split ring resonators~\cite{Pendry1999_ITMT}, many nanostructures have been demonstrated to exhibit similar features~\cite{Cai2010_book,Soukoulis2011_NP,Zhao2009_materialtoday,Zheludev2012_NM}. A lot of work has been done to study  structures that support magnetic resonances~\cite{Cai2010_book,Zheludev2012_NM,Zhao2009_materialtoday,Peng2007_PRL,Schuller2007_PRL,Vynck2009_PRL,Evlykuhin2010_PRB,Paniague2011_NJP,Garixia_etxarri2011_OE, Kuznetsov2012_SciRep,Evlyukhin2012_NL,Fan2012_NP,Dominguez2013_SR,Kuznetsov2013_arxiv}. However those responses  have been mostly employed to demonstrate negative, zero, or ultrahigh refractive indexes, and other related applications including cloaking, imaging and beam shaping~\cite{Cai2010_book,Soukoulis2011_NP,Zheludev2012_NM,Peng2007_PRL,Schuller2007_PRL,Vynck2009_PRL,Zhao2009_materialtoday,Paniague2011_NJP,
 Fan2012_NP,Dominguez2013_SR,Wu2013_arxiv,Moitra2013_NP}, which constitute the dominant mainstream of the field of metamaterials. Very limited attention has been paid to the topic of scattering shaping utilizing those magnetic resonances, especially in the optical spectral regime~\cite{Jin2010_IEEE,Gomez-Medina2011_JN}.

 After recent prediction of the unidirectional superscattering of core-shell nanospheres~\cite{Liu2012_ACSNANO}, and the experimental demonstration of magnetic dipoles (MDs) supported by spherical silicon nanoparticles in the visible and near-infrared spectral regime~\cite{Kuznetsov2012_SciRep,Evlyukhin2012_NL}, the original proposal by Kerker~\cite{Kerker1983_JOSA} and the concept of Huygens source in the antenna theory~\cite{Kerker1983_JOSA,Love1976_RS,Jin2010_IEEE} have been reinvestigated from the perspective of OMRs. Now the research direction of scattering shaping for nanoparticles utilizing those magnetic resonances are attracting surging attention, rendering it another major topic in the mainstream of the fields of metamaterials and nanoplasmonics.

Here we review the recent progress in this research direction, starting in Section $2$ with the demonstration of magnetic resonances optically-induced in various structures for different spectral regimes, and put more emphasis on those resonances supported by  all dielectric structures in the optical regime. Then in Section $3$ we summarize the series of works on scattering shaping for nanoparticles based on those magnetic resonances obtained, including the cases of interferences between spectrally separated (with different resonant wavelengths) and overlapped(with the same resonant wavelength)  ED and MD  responses, and also with other higher order modes. In Section $4$ we discuss Fano resonances involving OMRs. Section $5$ concludes with both summary and outlook. We note here that in this review we focus on scattering manipulation through OMRs of dielectric nanoparticles or the dielectric components of hybrid metal-dielectric nanoparticles. The already well known OMRs of metallic structures, such as split ring resonators, cut-wire pairs, fishnet structures and so on will not be the focus of this review.


\section{Optically-induced magnetic responses}

Since the seminal work of John Pendry on artificial magnetism of split ring resonators~\cite{Pendry1999_ITMT} and his reinvestigation on the Veselago Lens from the new perspective of perfect resolution~\cite{Pendry2000_PRL}, tremendous efforts have been made to search for structures to exhibit OMRs~\cite{Cai2010_book,Soukoulis2011_NP,Zheludev2012_NM}. Inspired by the work on split ring resonators, a lot of structures have been demonstrated to support OMRs based on the strong plasmonic responses of metallic elements, such as fishnet structures~\cite{Brueck2005_PRL,Valentine2008_nature}, coupled metal wires~\cite{Shalaev2005_OL,Dolling2005_OL}, and plasmonic nanoparticle clusters~\cite{Alu2009_OE5723,Shafiei2013_NNT}, to name but a few.  At the same time, the operating spectral regime has also being pushed from microwave all the way down to the visible range~\cite{Cai2010_book,Soukoulis2011_NP,Zheludev2012_NM}. It is worth mentioning that, though through design optimizations, the structures based on metallic inclusion can support OMRs in the optical regime~\cite{Cai2010_book,Soukoulis2011_NP,Zheludev2012_NM}, the high losses of metal elements in this spectral regime make those structures highly lossy and hinder severely many further possible applications and extensions. The losses also impose great limitations on the attempts to further shrink the wavelengths of the magnetic resonances.

The OMRs supported come from the circulating displacement currents and thus the metallic elements are not inevitably required.
To deal with the intrinsic limitations imposed by metallic elements, a lot of efforts have been made to search for OMRs in all-dielectric structures without any metallic inclusions~\cite{Zhao2009_materialtoday,Peng2007_PRL,Vynck2009_PRL,Schuller2007_PRL,Popa2008_PRL,Zhao2008_PRL027402,Brener2012_PRL}.
The investigations of all-dielectric OMRs started from the microwave regime~\cite{Zhao2009_materialtoday} and then the resonance wavelength has been further shrunk down to the mid-wavelength infrared regime~\cite{Brener2012_PRL}. The OMRs demonstrated are based on the localized resonances supported by high permittivity dielectric particles (cylinders, cubes or other shapes), which support strong circulating displacement currents in the transverse plane. Figure~\rpict{figure1}(b) shows both the ED and MD supported by cubic Tellurium resonators fabricated on the BaF$_2$ substrate [Fig.~\rpict{figure1}(a)]. The transmission and reflection spectra show specifically that  MD is supported at the wavelength of approximately $9~\mu$m, which is in the mid-wavelength infrared regime.

\pict[0.6]{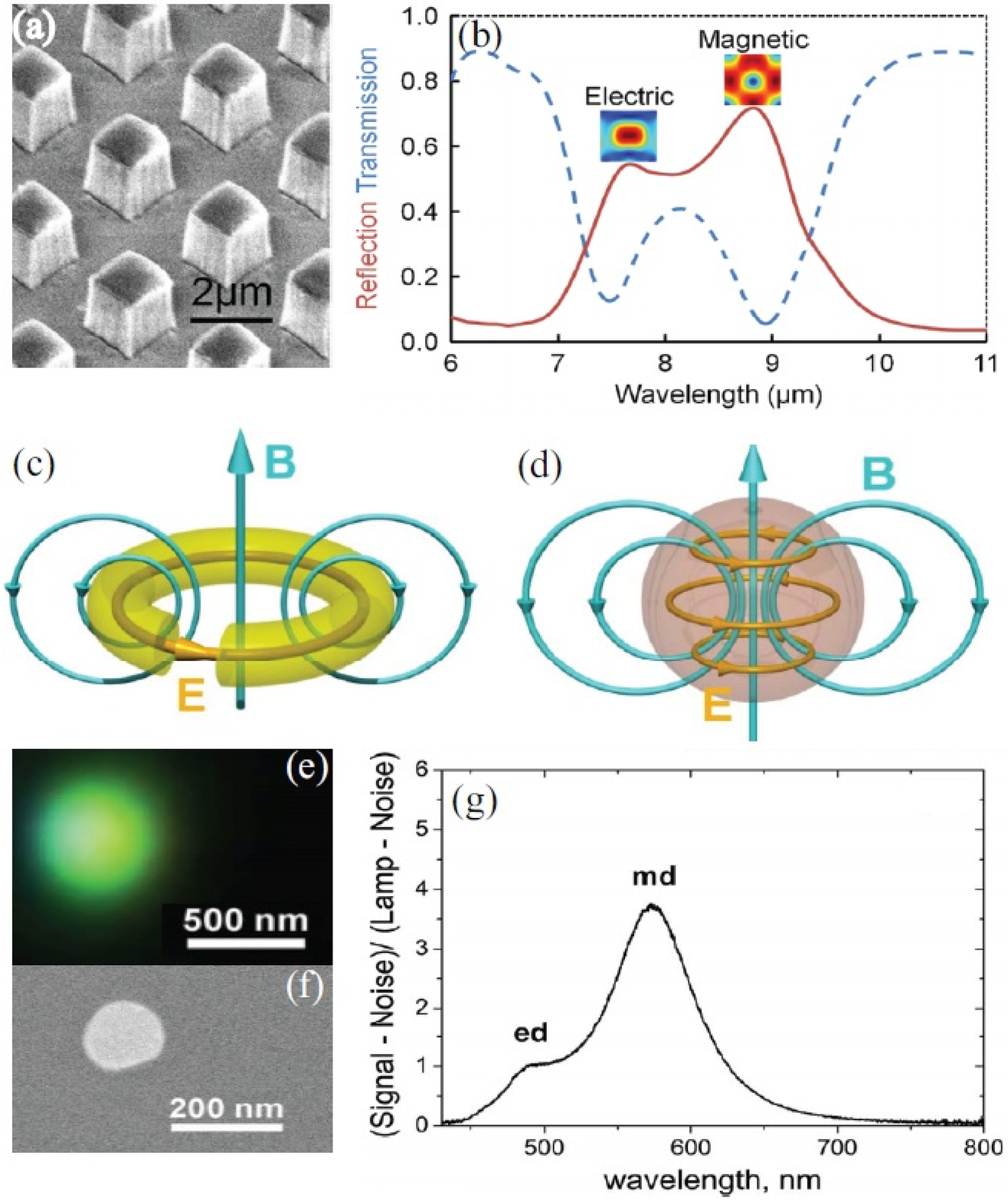}{figure1}{\small Examples of structures with optically-induced magnetic responses. (a) Fabricated arrays of Tellurium resonators on the substrate of BaF$_2$ and (b) the measured transmission (reflection) spectra with dips (peaks) induced by the electric and  magnetic dipoles supported by individual resonators. The insets show the electric field distribution in the transverse plane for the corresponding dipoles. (c) and (d) Schematic illustrations for the optically induced magnetic responses supported by a split-ring resonator and a silicon sphere, respectively. (e) Dark-field microscope, (f) SEM images, and (g) experimental dark-field scattering spectra of a single silicon sphere~\cite{Brener2012_PRL,Kuznetsov2012_SciRep}.}

The resonance wavelengths of the OMRs supported by all-dielectric structures can be further pushed down to the visible optical regime, as has been theoretically predicted by Evlykuhin \textit{et al.}~\cite{Evlykuhin2010_PRB} and Garc\'{i}a-Etxarri \textit{et al.}.~\cite{Garixia_etxarri2011_OE}.  They show that silicon spheres can directly support the  MDs and the resonance wavelengths are highly tunable through changing the radii of the silicon spheres. The MDs are the fundamental Mie resonances, which are cavity-type modes with most of the energy confined within the sphere. The origin of the magnetism is shown schematically in Fig.~\rpict{figure1}(d), with the split-ring resonator also shown for comparison in Fig.~\rpict{figure1}(c). For both resonators, the OMRs come from the strong circulating displacement currents (similar to the E field distribution), due to which the H fields are significantly enhanced.

Shortly after the theoretical predictions, breakthroughs have been made to demonstrate experimentally the existence of optically induced MDs supported by silicon spheres in both the visible and near-inferred spectral regimes~\cite{Kuznetsov2012_SciRep,Evlyukhin2012_NL}. Figures~\rpict{figure1}(e)-(g) show the silicon spheres fabricated, through the SEM [Fig.~\rpict{figure1}(f)] and dark-field microscope [Fig.~\rpict{figure1}(e)] images. Both ED and MD are visible in the  experimental dark-field scattering spectra of the single silicon sphere shown in Fig.~\rpict{figure1}(g). Other experimental work that accompanies or appears after the aforementioned demonstrations to show OMRs in the visible or near-infrared spectral regimes  are based on silicon structures, with optical or electron irradiation excitations~\cite{Shi2012_AM,Shi2013_NC,Albella_JPCC,Coenen2013_acsnano}.

\section{Scattering shaping utilizing magnetic responses}

It is well known that to introduce magnetic resonances into scattering systems brings extra freedom for the scattering engineering due to the interplay of both electric and magnetic responses, of which the Kerker's proposal of backward scattering suppression and forward scattering enhancement serves as an outstanding example~\cite{Kerker1983_JOSA}. Another related example is the concept of Huygens source in the antenna theory~\cite{Love1976_RS}. For both examples the scattering pattern shows two unusual features compared to that of an individual ED or MD: (1) suppressed backward scattering and enhanced forward scattering; and (2) azimuthally symmetric scattering. The features come directly from the interference of the ED and the MD, under the conditions of: (1) ED and MD overlaps with the same magnitude and (2) the two resonances are in phase.  When both conditions are satisfied, ED and MD can interfere totally destructive in the backward direction and constructively in the forward direction, leading to suppressed backward scattering and enhanced forward scattering. Both the proposal of Kerker and the concept of Huygens source were originally based on the magnetic materials.  However there are very limited kinds of materials that directly support magnetic responses, which at the same time can usually only operate in narrow spectral regimes, and are accompanied by high losses. That is why though a lot of related ideas on scattering shaping utilizing magnetic materials have been put forward~\cite{Nieto2010_OE,Garcia-Camara2010_OE},  unfortunately they have not attracted much attention.

Stimulated by the rapid progress of the fields of plasmonics and metamaterials, and especially the introduction of OMRs, it has been realized that OMRs might be employed for scattering shaping~\cite{Jin2010_IEEE,Gomez-Medina2011_JN}. But a great research activity in this direction still requires further stimulation. Such an impetus comes from the recent theoretical investigation on unidirectional superscattering of core-shell nanospheres and experimental demonstration of OMRs supported by silicon spheres in the visible spectral regime~\cite{Kuznetsov2012_SciRep,Evlyukhin2012_NL}. Since then, a lot of attention has been attracted to the reevaluation of Kerker's proposal and the concept of Huygens source from the perspective of OMRs.

\pict[0.9]{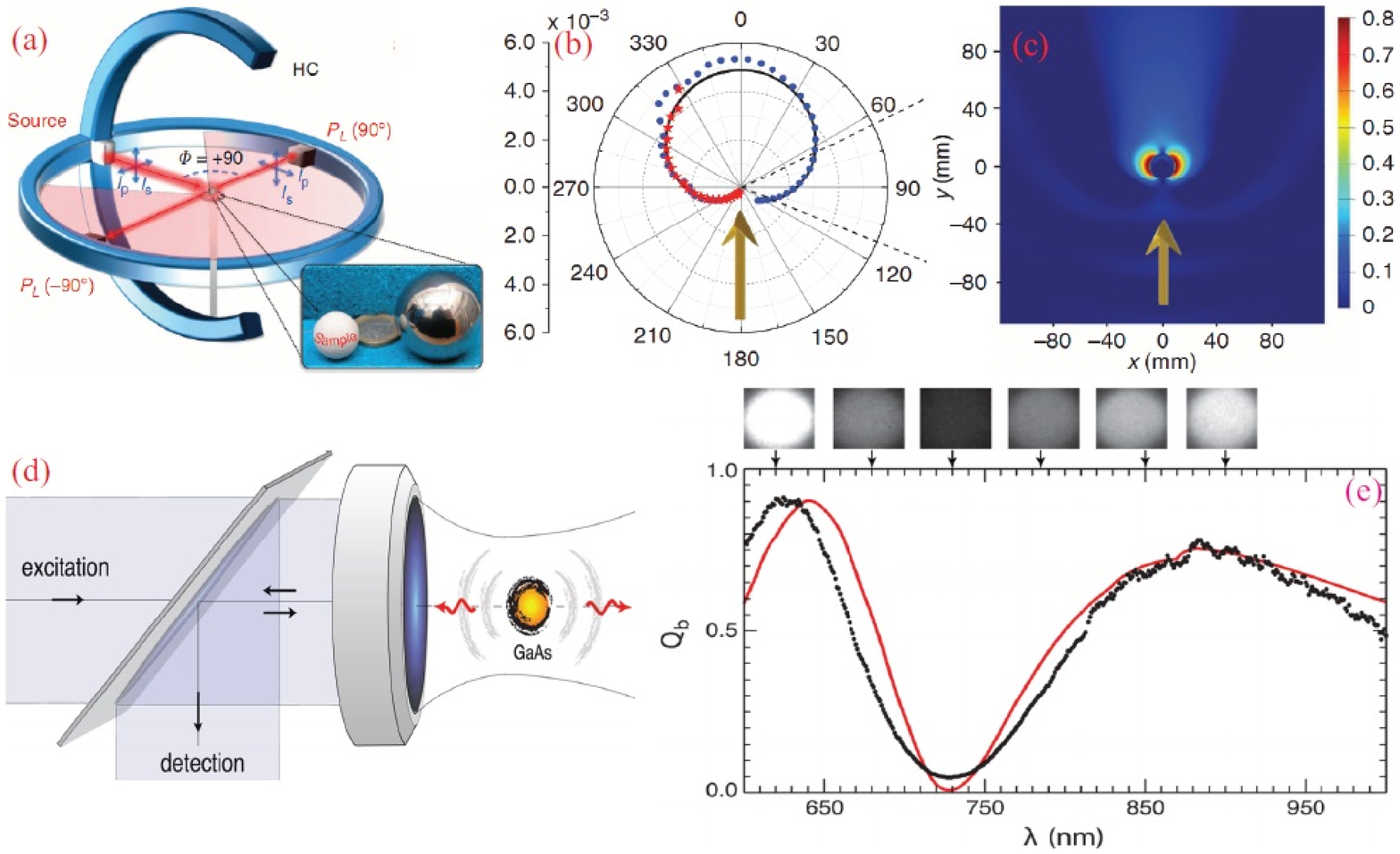}{figure2}{\small Measurements of the magnetic responses. (a) Schematic illustration of the experimental setup for the measurement of the scattering of the high permittivity dielectric particles in the GHz spectral regime. The sample put in the centre is highlighted on the side.  (b) and (c) show the far-field scattering pattern and the near-field intensity distributions respectively for the sample at the wavelength of approximately $83$ mm. (d) Schematic illustration of the experimental setup for the measurement of the backward scattering  of the GaAs pillar in the visible and near-infrared spectral regime. (e) Backward scattering spectrum (dotted: experimental results; solid: theoretical calculations). The images on top show the backward scattering images at different operating wavelengths~\cite{Geffrin2012_NC,Person2013_NL}.}

\subsection{Spectrally separated electric and magnetic dipoles with different resonant wavelengths}

 Fuelled by the experimental demonstration of OMRs with all-dielectric nanoparticles, related work on scattering shaping utilizing OMRs emerged and has been mostly performed on the all-dielectric platforms. The first idea that has been tested is the Kerker's proposal (Huygens source) with suppressed backward and enhanced forward direction scattering. This requires the presence of both ED and MD, which are in phase and of the same magnitude. Those conditions can be satisfied by single high permittivity dielectric particles, as both ED and MD are supported by such particles (see Fig.~\rpict{figure1}).

The demonstration was firstly done in the GHz regime by Geffrin \textit{et al.}~\cite{Geffrin2012_NC} and the results are partly shown in Figs.~\rpict{figure2}(a)-(c). Figure~\rpict{figure2}(a) shows schematically the experimental setup with the high permittivity dielectric particle put int he centre (for more details of the specific material used, see Ref.~\cite{Geffrin2012_NC}). Both the near-field intensity distribution [Fig.~\rpict{figure2}(c)] and far-field scattering pattern [Fig.~\rpict{figure2}(b)] indicate clearly the backward scattering suppression and forward scattering enhancement. The experimental demonstration in the visible and near-infrared spectral regime was done by Person \textit{et al.}~\cite{Person2013_NL} with GaAs nanoparticles, as shown in Figs.~\rpict{figure2}(d) and 2(e). According to the backward scattering spectrum [Fig.~\rpict{figure2}(e)], at the wavelength of approximately $730$~nm the backward scattering has been totally suppressed, where ED and MD are both present and interfere destructively to cancel the backward direction. The limitation of this work by Person \textit{et al.} is that the forward scattering has not been directly measured. A more comprehensive demonstration in the visible spectral regime has been done by Fu \textit{et al.}~\cite{Fu2013_NC} with silicon nanoparticles. The experimental setup is shown schematically in Fig.~\rpict{figure3}(a), where both forward and backward scattering are measured. Direct CCD (charge-coupled device) images of silicon nanoparticles of different sizes are shown in Figs.~\rpict{figure3}(b) and 3(c), which correspond to backward and forward scattering, respectively.

 At the same time, the idea of scattering shaping through inferencing ED and MD has also been applied to dipole sources~\cite{Filonov2012_APL,Krasnok2012_OE,Rolly2012_OE} and to the demonstration of optically induced antiferromagnetism in hybrid metamaterials~\cite{Miroshnichenko2012_ACSNANO}.

We note here that Kerker \textit{ et al.} also proposed another case that both ED and MD are present of the same magnitude but are out of phase~\cite{Kerker1983_JOSA}. For this case, in contract to the case of ED and MD in phase, the forward scattering is suppressed and backward scattering is enhanced. This case has also been demonstrated in Refs.~\cite{Geffrin2012_NC,Fu2013_NC}.  However it should be kept in mind that according to the optical theorem~\cite{Bohren1983_book,Alu2010_JN}, for the plane wave incidence
the forward scattering is usually approximately proportional to the overall scattering. For this case of suppressed forward scattering, although the backward scattering is relatively enhanced, the overall scattering is small, rendering this case not as promising as the case of two dipoles in phase. Also according to optical theorem, the case of  forward scattering cancellation and enhanced backward scattering is not possible for plane wave incidence.

\pict[0.9]{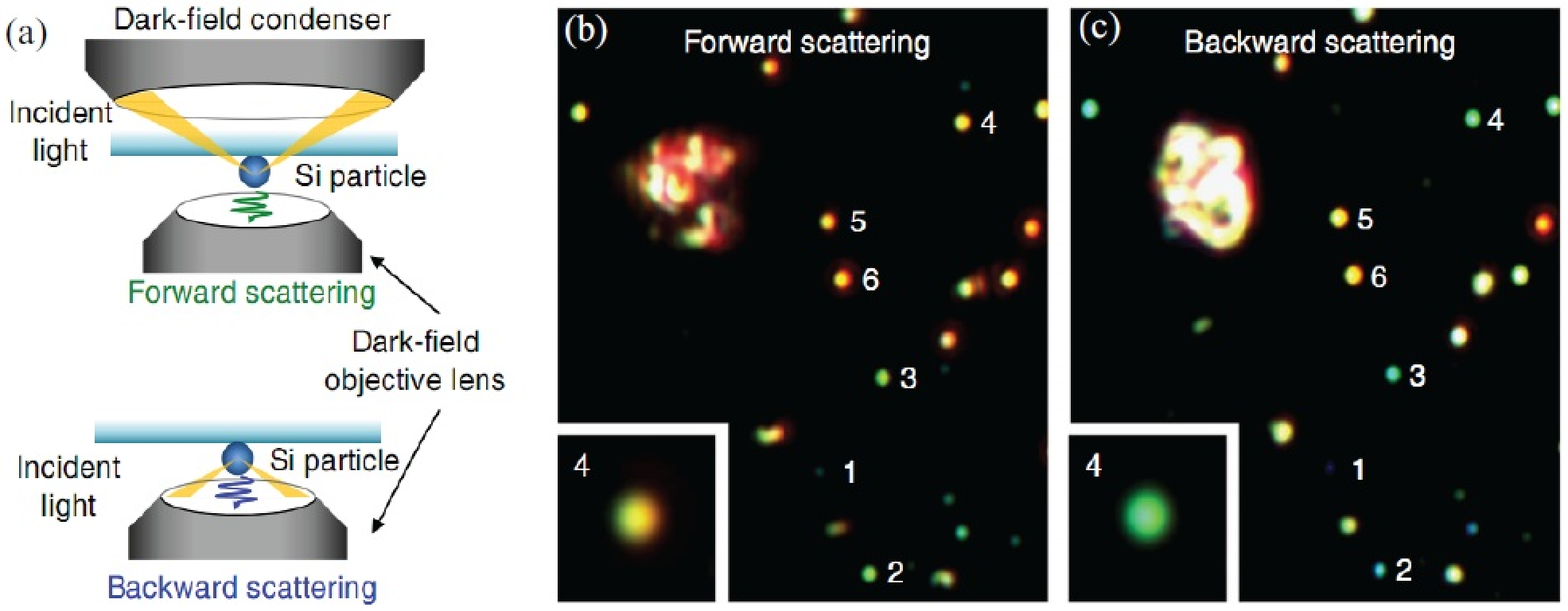}{figure3}{\small  (a) Schematic illustration of the experimental setup for the measurement of the scattering (both in the forward and backward directions) of the silicon particles in the visible spectral regime. (b) and (c) show the CCD images of true colors for the forward and backward direction scattering respectively of the silicon particles with diameters ranging from $100$ to $200$ nm. The enlarged dark-field microscope image of silicon particle No. $4$ are shown in the corners of (b) and (c)~\cite{Fu2013_NC}.}

\subsection{Spectrally overlapping electric and magnetic dipoles with the same resonant wavelength}

The demonstrations discussed in the subsection above are mostly based on individual homogenous dielectric particles, for which though both ED and MD can be supported, they are usually separated spectrally  with different resonant wavelengths [see Figs.~\rpict{figure1}(b) and 1(g) for example].  For those particles, spectral points or regimes of interest can still be found where scattering can be efficiently controlled. However, those spectral points or regimes of interest are not in the resonant region and consequently the total scattering is inevitably relatively small. This feature imposes the limitations on further applications that require large scattering cross sections.

\pict[0.9]{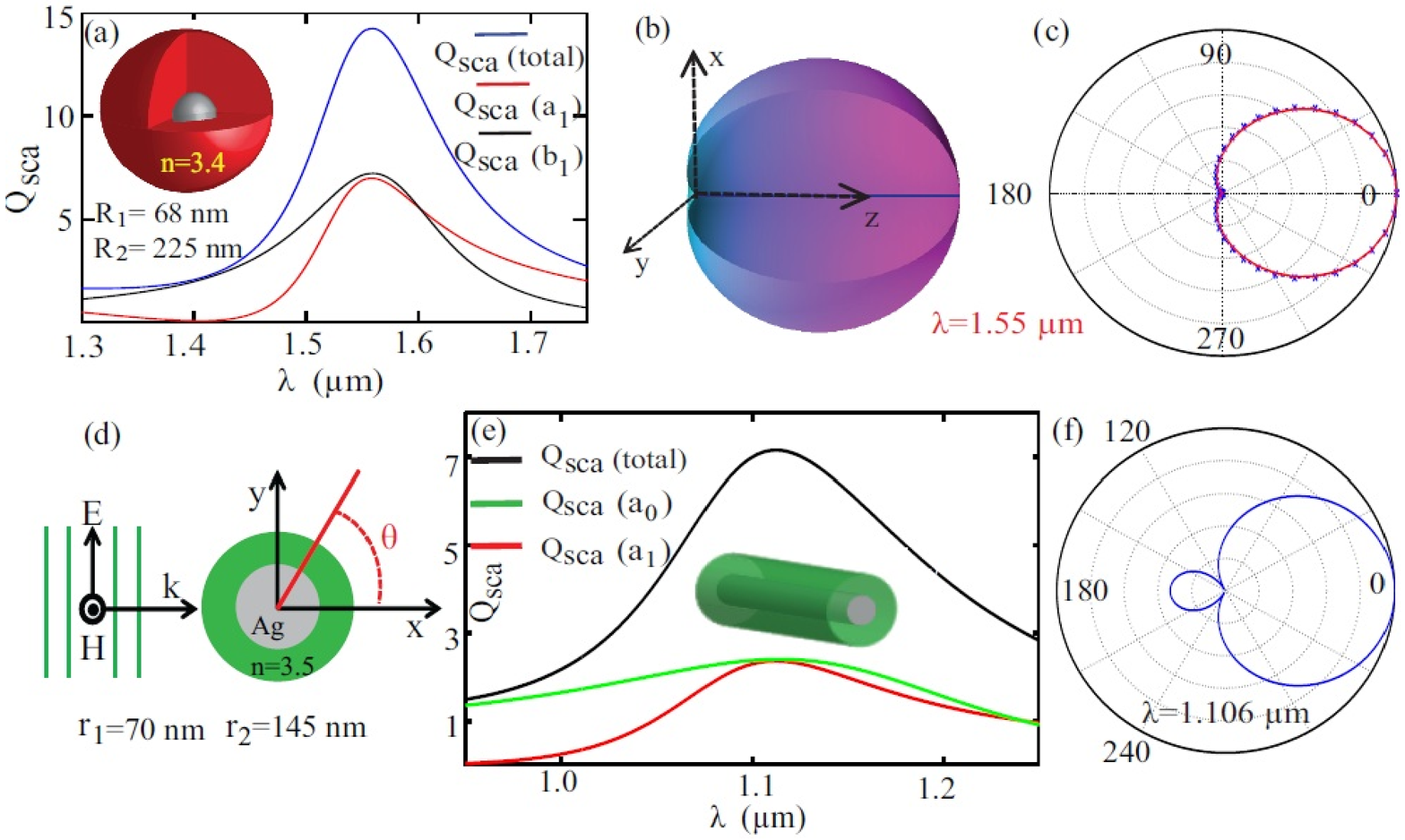}{figure4}{\small  (a) Scattering spectrum of core-shell nanoparticles. Both the contributions from ED ($a_1$) and MD ($b_1$) are shown. (b) and (c) show respectively the 3D  and 2D far-field scattering patterns at the overlapping resonant wavelength of ED and MD at $1550$~nm. The incident plane wave is propagating along $z$ direction and in (c) solid line and line with crosses correspond to patterns on the $x-z$ and $y-z$ planes respectively. (d) The scattering configuration and (e) scattering spectrum of the core-shell nanowire. Both the contributions from MD ($a_0$) and ED ($a_1$, single angular momentum channel) are shown. (c) shows the far-field scattering pattern at the overlapping resonant wavelength of ED and MD at $1106$~nm~\cite{Liu2012_ACSNANO,Liu2013_OL2621}.}

To enhance the scattering cross sections of nanoparticles, recently the concept of superscattering has been put forward by Ruan \textit{et al.}~\cite{Ruan2010_PRL} and the basic approach is to spectrally overlap different modes at the same resonant wavelength. To realize this the first problem that has to be solved is to avoid the near-field strong coupling of the modes involved, or supermodes will be formed and spectrally they will be separated with different resonant wavelengths. A direct way to avoid near-field mode coupling is to employ the different eigenmodes of the nanoparticles, which are orthogonal and there is no coupling between them. But for an individual particle, eigenmodes usually  have different resonant wavelengths, and shift those wavelengths to the same point would involve complicated structure engineering and usually high losses of the structure would be induced~\cite{Ruan2010_PRL}.

A natural question arises that whether it is possible to overlap ED and MD at the same resonant wavelength. This is highly desirable as this would simultaneously lead to superscattering and efficient scattering shaping. As we mentioned before, the  MDs supported by dielectric spheres are cavity-type modes, for which the resonant wavelength have to approximately satisfy  the requirement that the phase accumulation along the diameter is $2\pi$~\cite{Garixia_etxarri2011_OE,Kuznetsov2012_SciRep}. At the same time, it is well known that the EDs of metallic particles are surface-type modes, for which the resonant wavelength can be easily decided through applying the Bohr condition~\cite{Yang2012_PT,Liu2013_thesis}. Based on those understandings, it is demonstrated that in core-shell metal-dielectric nanoparticles, ED and MD can be engineered to overlap at the same resonant wavelength~\cite{Liu2012_ACSNANO,Paniague2011_NJP}, with the same magnitude and the two dipoles in phase. Similar to the demonstrations mentioned in the subsection above, the suppressed backward scattering and enhanced forward scattering is achieved. However, the difference for core-shell nanoparticles compared with those homogenous dielectric nanoparticles is that the scattering shaping spectral region is in the superscattering regime, accompanied by enhanced overall scattering cross sections~\cite{Liu2013_thesis,Liu2012_ACSNANO}. The scattering spectrum of the core-shell nanoparticle is shown in Fig.~\rpict{figure4}(a), which indicates that ED ($a_1$) and MD ($b_1$) are overlapped with the same resonant wavelength of approximately $1550$~nm. The scattering patterns [3D in Fig.~\rpict{figure4}(b) and 2D Fig.~\rpict{figure4}(c)] at this wavelength show obvious features of backward scattering suppression and forward scattering enhancement. At the same time, the total scattering efficiency at this point is almost twice of that of the single channel limit, indicating the existence of superscattering and low loss~\cite{Ruan2010_PRL}. It has been further demonstrated that to align such core-shell nanoparticles in an array, the directionality of the scattering pattern can be further improved and the unidirectional scattering pattern is an effectively broadband response~\cite{Liu2012_ACSNANO}.

Similar ideas can also be applied to the scattering problem of core-shell nanowire as shown in Figs.~\rpict{figure4}(d)-(f)~\cite{Liu2013_OL2621}. Similar to the case of core-shell nanosphere, ED and MD can be made to overlap with the same resonant wavelength of $1106$~nm [Fig.~\rpict{figure4}(e)] and thus the scattering is at the superscattering regime. The difference is that the curve denoted by $a_1$ in Fig.~\rpict{figure4}(e) correspond to only one of the two degenerate EDs with opposite angular momentum. That is to say at the resonant wavelength, the magnitude of the ED is twice of that of the MD (corresponding to the $a_0$ line). The scattering pattern of this case is shown in Fig.~\rpict{figure4}(f). In contrast to the case of core-shell nanosphere, the scattering is suppressed not at the backward direction but at a pair of angles that satisfy $\alpha \cos(\theta)+1=0$, where $\theta$ is the scattering angle [Fig.~\rpict{figure4}(d)] and $\alpha$ is the ratio of the magnitude of ED to that of MD, with $\alpha=2$ in this case. Those results generalize the proposal of Kerker and the concept of Huygens source, which originally put the emphasis only on the scattering control in the forward and backward directions. It is worth mentioning that in Figs.~\rpict{figure4}(d)-(f) we show only $p$ polarized incidence wave, and actually the scattering pattern is dramatically different for $s$ polarized incident wave~\cite{Liu2013_OL2621}. Based on this high polarization-dependence, the core-shell nanowires could possibly be used as polarization splitters or filters and for other related applications.

Up to now, we have discussed only the approach to overlap ED and MD with the same resonant wavelength in hybrid metal dielectric structures, where EDs come from the plasmonic responses of metallic elements and MDs come from the displacement currents in high permittivity dielectric structure. It is recently also demonstrated that single homogenous silicon nanodisks can support such ED and MD pairs with the same resonant wavelength~\cite{Staude2013_acsnano}. The magnitude ratio of magnitude of ED to that of MD is neither $1$ (as for core-shell nanospheres) nor $2$ (as for core-shell nanowires), indicating that the scattering suppression angle would be different from what have already been demonstrated above. However, this work by Staude \textit{et al.}~\cite{Staude2013_acsnano} has investigated experimentally only arrays of such silicon particles. The corresponding angular scattering patterns and and total scattering intensities of individual disks are still to be identified.

\subsection{Scattering shaping with higher-order electric and magnetic modes}

As has been discussed above, the interference of ED and MD offers high flexibilities for various scattering shaping. The dipole approximations can be applied when the magnitudes of the dipoles are overwhelmingly larger than those of the other higher order modes. But at some spectral regimes, the dipole approximation does not hold any more and the contributions from higher orders modes have to be considered~\cite{Evlyukhin2011_PRB,Liu2013_OL2621,Krasnok2013_arxiv,Rolly2013_arxiv}. Figure~\rpict{figure5}(a) shows the scattering spectrum of the same nanowire as shown in Fig.~\rpict{figure4}(d), but in another spectral regime. At this regime, the contribution from the ED ($a_1$) is negligible, and the MD ($a_0$) and  electric quadrupole (EQ) are dominant. The scattering patterns in Fig.~\rpict{figure5}(b) and (c) come from the interference of the MD and EQ. The central resonant wavelength of EQ is $887$ nm and the phase of the EQ would change sign at this point, leading to drastically different scattering patterns at the wavelengths on the different sides of this wavelength point, as indicated by Fig.~\rpict{figure5}(b) and (c)~\cite{Liu2013_OL2621}.

\pict[0.9]{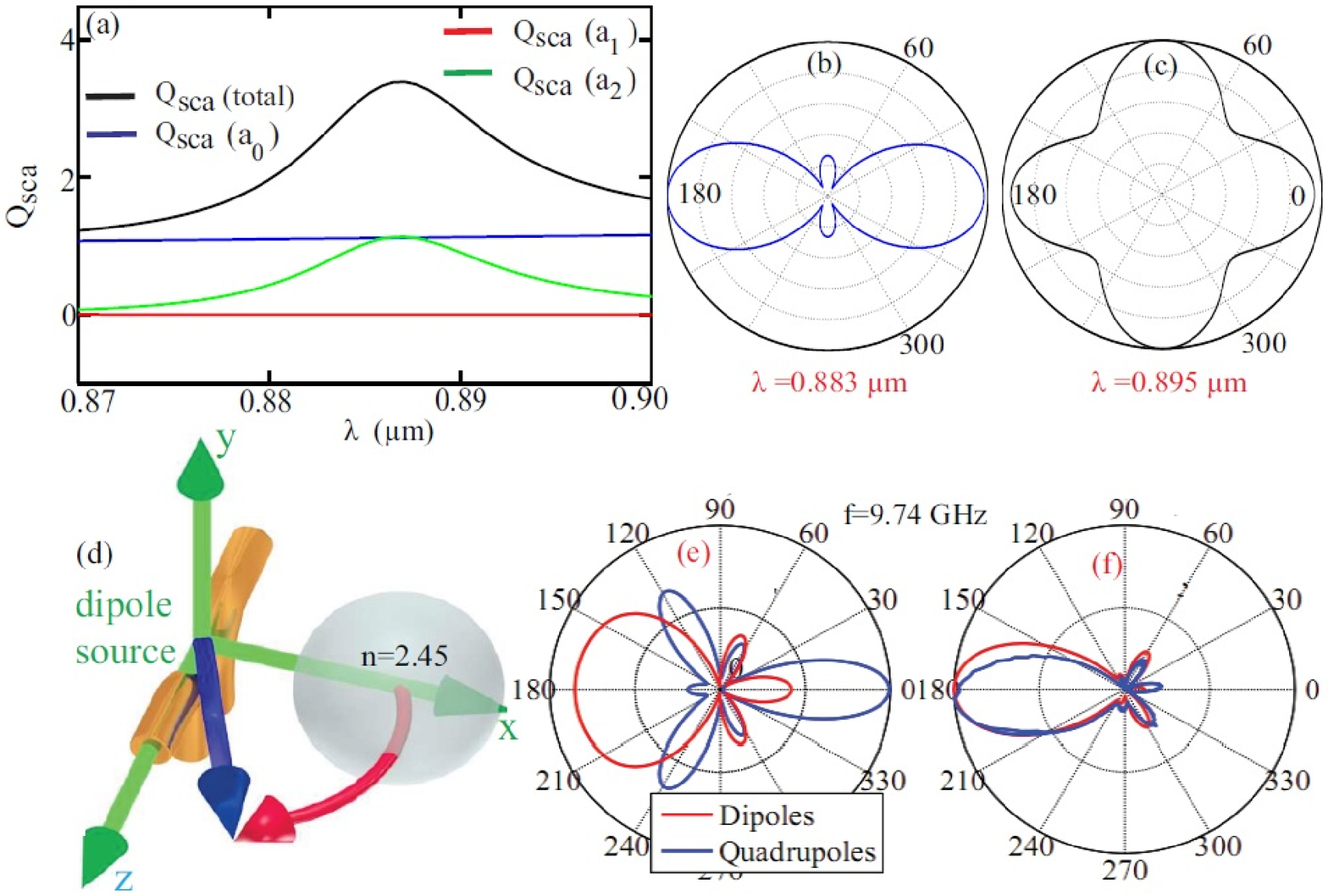}{figure5}{\small  (a) Scattering spectrum of the core-shell nanowire as shown in Fig.~\rpict{figure4}(d). The contributions from ED ($a_1$), MD ($b_1$) and EQ ($a_2$) are shown. (b) and (c) show respectively the scattering patterns at the wavelengths of  $883$~nm and $895$~nm respectively.  (d) shows the experimental setup for the dipole source emission shaping through a dielectric ($n=2.45$) sphere. (e) shows the the theoretical calculations of the scattering pattern in the $x-z$  plane at the frequency of $f=9.74$ GHz when only dipoles (red curve) or quadrupole (blue curve) have been taken into consideration. (d) Red curve:theoretical results of the scattering pattern with $20$ multipoles considered. Blue curve: the experimentally measured results. For (d)-(f) the distance between the emitter and the dielectric particle is $10$ mm~\cite{Liu2013_OL2621,Rolly2013_arxiv}.}

The higher order modes are usually present when it is dipole source excitation, and thus the contributions from higher order modes usually have to be considered when the OMRs are employed to shaping the emission of dipole sources~\cite{Krasnok2013_arxiv,Rolly2013_arxiv}. Figure~\rpict{figure5}(d) shows schematically the configuration for dipole emission shaping using high permittivity dielectric spheres. The theoretical calculations for the scattering pattern in the $x-z$  plane at $f=9.74$ GHz (emitter-to-particle gap is $10$ mm) are shown in Fig.~\rpict{figure5}(e) when only dipoles or only quadrupoles are considered. The measured results are shown in Fig.~\rpict{figure5}(f) by the blue cure. Further theoretical calculations have been preformed with $20$ multipoles and the results are shown in Fig.~\rpict{figure5}(f) by the red cure. It is clear from Fig.~\rpict{figure5}(e) and (f) that higher order modes contribute significantly to the directivity of the scattering pattern. The limitation of the work by Rolly \textit{et al.} is that they investigate only the case of separated dipole source and dielectric particle antenna, and the specific contributions from higher order modes are not clear~\cite{Rolly2013_arxiv}.  A more comprehensive study has been conducted by Krasnok \textit{et al.}, where it is shown that introducing an extra notch would further improve the directivity of the dipole emission, and the detailed contributions from higher order modes have been clarified from the perspectives of both magnitude and phase~\cite{Krasnok2013_arxiv}.

\section{Fano resonances involving optically-induced magnetic responses}

In contrast to the Lorentz resonance, the line-shape of Fano resonance is intrinsically asymmetric, which is induced by the interference of a broad spectral line background state and a narrow discrete state~\cite{Fano1961_PR,Miroshnichenko2010_RMP,Lukyanchuk2011_NM}.
Recently stimulated by the fields of plasmonics and metamaterials, a lot of investigations on Fano resonances have been conducted in plasmonic nanostructures~\cite{Miroshnichenko2010_RMP,Lukyanchuk2011_NM,Hopkins2013_nanoscale,Hopkins2013_PRA,Rybin2013_PRB}. Due to the same reason that we have mentioned above that most materials have only dominant electric resonances, the Fano resonances studied usually involve the interference of only electric modes. Inspired by the recent rapid progress of OMRs related studies, it is demonstrated that OMRs can be employed for Fano resonance tuning, with the magnetic modes serving as the broad background state or  narrow discrete state~\cite{Ruan2013_JPCC7324,Liu2013_OL2621,Liu2012_PRB081407,Miroshnichenko2012_NL6459,Chen2012_small1503,Yang2013_APL111115,Wu2013_arxiv}.

\pict[0.9]{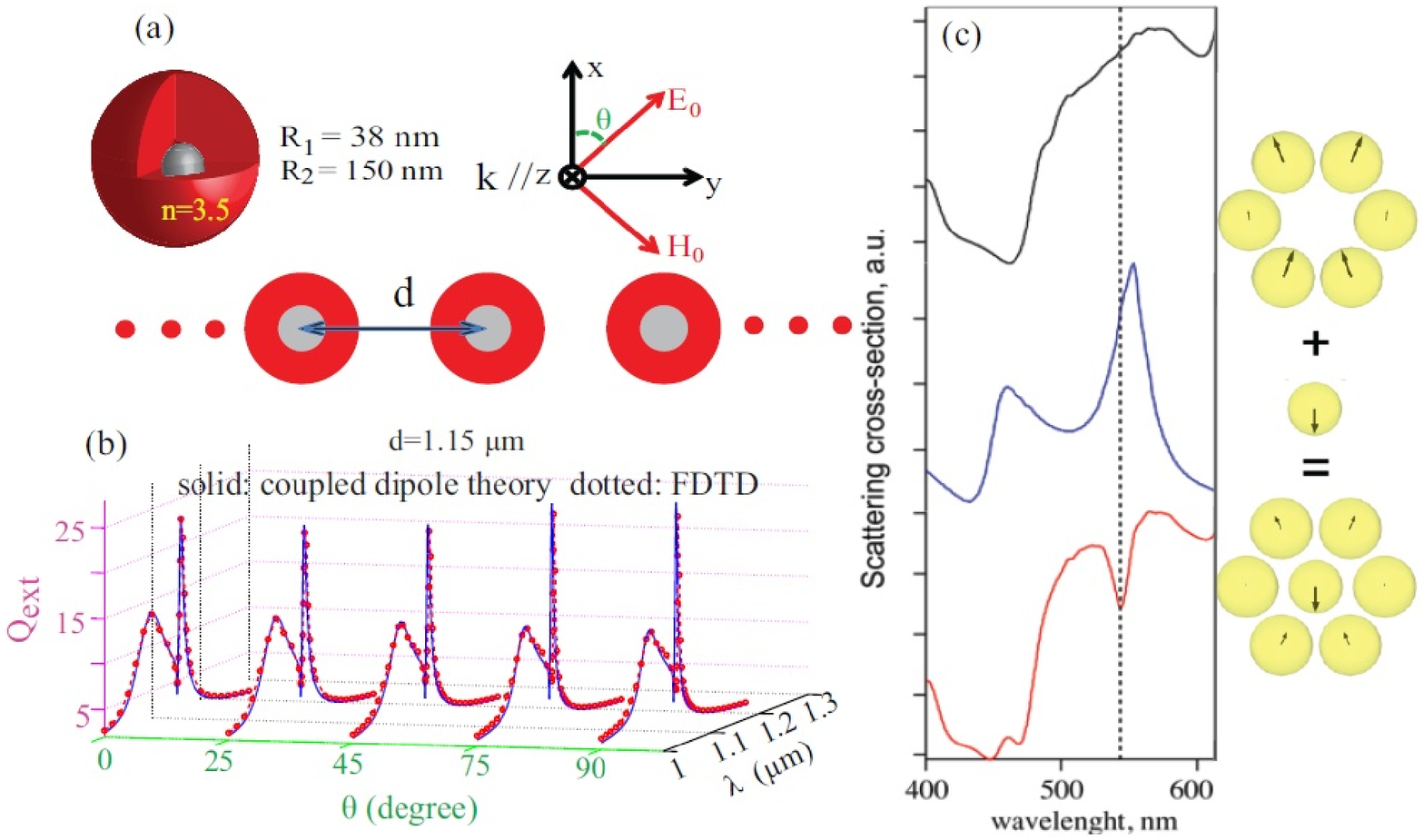}{figure6}{\small  (a) Scattering of an incident plane wave by a 1D array of core-shell silver-dielectric nanoparticles. The particle is similar to that shown in Fig.~\rpict{figure4}(a) with different parameters, result in a overlapping resonant wavelength of $1096$ nm. The polarization direction is indicated by $\theta$. (b) The corresponding scattering spectra of such an array for different $\theta$. (c) Scattering spectral for different silicon based structures: sphere haxamer (top, $R=75$~nm and inter-particle gap is $10$ nm), single silicon sphere (middle, $R=65$~nm) and the heptamer (bottom) combined from the haxamer and the single silicon sphere~\cite{Liu2012_PRB081407,Miroshnichenko2012_NL6459}.}

Fano resonances that involve both electric and magnetic modes can show unique features that are not exhibited by those involving only electric responses~\cite{Liu2012_PRB081407}.  Figure~\rpict{figure6}(a) shows the configuration of light scattering by a 1D array of core-shell nanoparticles. Similar to what is shown in Fig.~\rpict{figure4}(a), the nanoparticle is tuned to support both ED and MD, which are engineered to resonate at the some wavelength of approximately $1096$ nm. It is known that this kind of array support Fano resonances, which comes from the interference of the broad resonances of individual particles and the narrow resonance of the whole array~\cite{Markel2005_JPBMO}. If the individual particle support only dominantly ED or MD, as for metallic particles or higher permittivity dielectric particles in some specific spectral regimes, the Fano resonances obtained would be highly polarization dependent~\cite{Liu2012_PRB081407,Markel2005_JPBMO}. However, if the individual particle is the core-shell nanoparticle as shown in Fig.~\rpict{figure6}(a), the two resonant modes supported can interfere simultaneously with the narrow resonance of the periodic array through diffractive coupling, producing polarization independent Fano resonances, despite the fact that such periodic structures do not possess azimuthal symmetry [see Fig.~\rpict{figure6}(b)].

Fano resonance can also be induced by solely OMRs~\cite{Miroshnichenko2012_NL6459}. Figure~\rpict{figure6}(c) shows such a Fano resonance [bottom of Fig.~\rpict{figure6}(c)], which comes from the interference of the broad background state [the collective response of the silicon haxamer, top of Fig.~\rpict{figure6}(c)] and the narrow discrete state [MD supported by a single silicon particle, middle of Fig.~\rpict{figure6}(c)]. Metallic nanoparticles arranged in a similar way can also support Fano resonances~\cite{Miroshnichenko2012_NL6459}. However as we have mentioned above, magnetic modes of silicon spheres are cavity-type modes with most of the energy confined within the particle. Consequently,  the Fano resonances formed in silicon heptamers are almost independent of the inter-particle gap distances, which is dramatically different the heptamers consisting of metallic nanoparticles.~\cite{Miroshnichenko2012_NL6459}

Related studies have also been conducted for single nanoparticles~\cite{Miroshnichenko2010_RMP,Lukyanchuk2011_NM,Ruan2013_JPCC7324,Liu2013_OL2621,Chen2012_small1503,Yang2013_APL111115, Zayats2013_OE8426}. Actually the dramatic change between the scattering patterns shown in Fig.~\rpict{figure5}(b) and (c) originates from the Fano resonance produced by the interference of MD (broad background state) with EQ (narrow state). Other (both concentric and non-centric) core-shell nanoparticles have also been studied and are demonstrated to support similar types of Fano resonances~\cite{Ruan2013_JPCC7324,Chen2012_small1503,Yang2013_APL111115,Zayats2013_OE8426}.


\section{Conclusion and outlook}

We have summarized briefly the recent progress on the scattering control of nanoparticles with OMRs. First, we have discussed the origin of the optically-induced magnetic modes supported by various nanostructures in different spectral regimes, with more emphasis on the recent breakthroughs on the magnetic modes supported by high-permittivity dielectric structures in the visible and near-infrared spectral regimes. Then we have reviewed the studies on the scattering shaping of nanoparticles utilizing those magnetic resonances, including the cases of both resonantly separated and
overlapped electric and magnetic dipoles, and also the cases involving higher-order electric and magnetic modes.
In addition, we have discussed the scattering control by Fano resonances involving optically-induced magnetic modes, which are supported by both individual nanoparticles and nanoparticle clusters.

Despite the rapid progress and various achievements in the field of scattering control with optically-induced magnetic resonances, there are still several immediate challenges laying ahead. Most studies have been done under the plane wave or dipole wave incidence, while the cases of other specially engineered incident waves, especially those carrying orbital and spin angular momentum~\cite{Berry1987_JMP,Padgett2009_LPR,Gabriel2012_OE}, have rarely been studied from the perspective of optical magnetism. At the same time, for other types of incident waves, a lot of higher order modes will be present. Although recently there appeared some related studies~\cite{Rolly2013_arxiv,Krasnok2013_arxiv}, the problem of how different modes contribute to the directivity and how to control the magnitude and phase of those modes have not been completely investigated. Other challenges include the study on magnetic modes supported by structures with gain and nonlinearity, and the integration of nanoparticles with optically-induced magnetic modes within optical circuits.

It is expected that other fundamental breakthroughs in the topic of light control utilizing magnetic resonances
 might be made from its merging with current vibrant fields, including spinoptics~\cite{Berry1987_JMP,Bliokh2008_NP,Yin2013_science,Hosten2008_science,
Xiao2010_RMP,Miroshnichenko2013_science}, graphene~\cite{Geim2007_NM,Grigorenko2012_NP,Xiao2010_RMP}, and topological insulators~\cite{Qi2011_RMP,Hasan2010_RMP,Khanikaev2013_NM,Rechtsman2013_nature}, which might help to find deeper space and extra dimensions to establish new physical principles, to demonstrate new phenomena in the field of nanophotonics, and to find numerous practical applications in sensing, imaging, nanoantennas, photovoltaic devices and so on.

\section*{Acknowledgements}

We are deeply indebted to many of our colleagues and collaborators, mostly to D. N. Neshev, O. Hess, and R. F. Oulton
for many useful discussions and suggestions. 


\providecommand{\newblock}{}

\end{document}